\documentclass[12pt]{article}
\usepackage[super]{natbib}
\usepackage{graphicx}
\usepackage{epstopdf, epsfig}
\usepackage{url}

\usepackage{footnote}
\usepackage{footmisc}
\makesavenoteenv{tabular}
\makesavenoteenv{table}

\renewcommand{\thefootnote}{\alph{footnote}}

\newcommand{\astfootnote}[1]{
\let\oldthefootnote=\thefootnote
\setcounter{footnote}{0}
\renewcommand{\thefootnote}{\fnsymbol{footnote}}
\footnote{#1}
\let\thefootnote=\oldthefootnote
}

\advance\hoffset by 8mm
\advance\voffset by -15mm \evensidemargin 1pt \oddsidemargin 1pt

\title{The fractal time growth of COVID-19 pandemic: an accurate self-similar model, and urgent conclusions}

\author{Alfonso M. Ga\~n\'an-Calvo$^{1,*}$ and Juan A. Hern{\'a}ndez Ramos$^2$\\
$^1$ Depto. de Ingenier\'{\i}a Aeroespacial y Mec\'anica de Fluidos,\\
 ETSI, Universidad de Sevilla, E-41092 Sevilla, Spain.\\
$^2$ Depto. de Matem{\'a}tica Aplicada a la Ingenier{\'\i}a Aeroespacial,\\
 ETSIAE, Universidad Polit{\'e}cnica de Madrid, 28040 Madrid, Spain.\\
$^*$amgc@us.es}

\begin{document}
\flushbottom
\maketitle
\thispagestyle{empty}

\begin{abstract}
Current available data of the worldwide impact of the COVID-19 pandemic has been analyzed using dimensional analysis and self-similarity hypotheses. We show that the time series of infected population and deaths of the most impacted and unprepared countries exhibits an asymptotic power law behavior, compatible with the propagation of a signal in a fractal network. We propose a model which predicts an asymptotically self-similar expansion of deaths in time before containment, and the final death toll under total containment measures, as a function of the delay in taking those measures after the expansion is observed. The physics of the model resembles the expansion of a flame in a homogeneous domain with a fractal dimension 3.75. After containment measures are taken, the natural fractal structure of the network is drastically altered and a secondary evolution is observed. This evolution, akin to the homogeneous combustion in a static isolated enclosure with a final quenching, has a characteristic time of 20.1 days, according to available data of the pandemic behavior in China. The proposed model is remarkably consistent with available data, which supports the simplifying hypotheses made in the model. A universal formulation for a quarantine as a function of that delay is also proposed.
\end{abstract}

\newpage
\section*{Introduction and proposal}
SARS-CoV-2 is a novel virus unfortunately affecting human species in an unprecedented way. The disaggregated time series accrued in Github by Ryan Lau (see \url{https://github.com/CSSEGISandData/COVID-19/tree/master/csse_covid_19_data/csse_covid_19_time_series})
for ``Confirmed'' $C(t)$ and ``Deaths'' $D(t)$ in 238 different countries and regions exhibit a wealth of information. These series for each country or geographical region could be considered a extremely useful disaggregated database to gauge the real magnitude of the pandemic. We have performed a global analysis of data made available up to March 26th, 2020, using the concept of self-similarity. We hypothesize that the expansion of the infection rapidly reaches an asymptotic homogeneous behavior once a minimum critical number of confirmed cases is known and a rapid expansion immediately follows. In this expansion, the virus exploits the inter-individual network of physical interactions among humans living in cities. This network is sufficiently large and populated to exhibit the drastically reduced degrees of freedom of the laws of big numbers: we observe that the data of those regions where no measures were enacted collapse into a common universal behavior. A self-similar simple law is found: the behavior is reduced to a universal time-power law of the type $\varphi\sim \tau^\alpha$ for the behavior of COVID-19 pandemic before containment measures are enacted, where $\tau$ is the appropriate non-dimensional time since the onset of the free expansion and $\varphi$ the non-dimensional time descriptor of the infected population and mortality. After containment measures are enacted, an exponential decay of mortality can be observed.

In this work, we propose a simple function showing an excellent fitting to both $C(t)$ and $D(t)$ series, which serves as a powerful predictor of the expansion of the pandemic before measures are taken: $\varphi=\left(\log\left(\beta +\exp(\tau)\right)\right)^\alpha$, where $\alpha$ and $\beta$ reflect the general expansion of the pandemic and the behavior at early times. $\tau=(t-t_0)/T_L$, where $T_L$ is the average infectivity time observed, about 20.1 days for this disease (which is calculated in this work from measurements), and $t_0$ the time where the expansion takes place, specific for each geographic region. We observe that $t_0$ is different for $C(t)$ and $D(t)$. From now on and for simplicity, we call $t_c$ and $t_d$ to the values of $t_0$ in series $C(t)$ or $D(t)$, respectively. The predictor is defined as $\varphi_c=C/C_C$ or $\varphi_d=D/(m C_C)$ for confirmed and deaths, where $C_C\simeq 1,2\times 10^4$ is a characteristic size of the pandemic infectious population. Here, $m$ is an average early mortality descriptor observed with respect to confirmed cases for this disease, which may depend on the population structure and health system. Interestingly, this factor exhibits a relative homogeneity around the value $m=0.15$, independently of the country, for the sixteen countries with the current highest impact of COVID-19, with the exception of three countries (Spain, France and Iran). In reality, $m$ gauges the initial mortality observed (or, properly speaking, {\it reported}) during the free expansion (before containment measures) of the pandemic, i.e. the average measured daily ratios $D(t)/C(t)$ during the initial self-similar expansion.

Our proposed function $\varphi$ exhibits a very early exponential evolution compatible with other proposed models (e.g. SIR and SEIR models\cite{Liu2016,Biswas2020,Iwata2020}), but rapidly approaches a power law as $\tau^\alpha$, where $\alpha$ is a fitting parameter with a value $\alpha\simeq 3.75$. This parameter represents the fractal dimension of a homogeneous network formed by the actual physical interactions among humans. It gauges the hidden depth of those interactions and the share of viable biological material among humans.

This work concludes with urgent recommendations according to our model and fittings to existing data. First, we strongly recommend to take a total confinement quarantine time or a complete selective isolation (including the whole network of potential contacts of the infected cases) {\it before} the first 100 cases of any unknown infection produced by a coronavirus are reported. If that is not possible, we recommend to take a quarantine of the order of two times the period given by $T_Q = T_L\exp\left((t_{measures}-t_c)/T_L\right)$, with $t_{measures}$ is the day when measures are enacted. COVID-19 exhibits a characteristic infection time $T_L=20.1$ according to our measurements from the evolution of the pandemic in China. Second, we strongly recommend to immediately size a health system to handle a potential death toll; in the case of COVID-19, according to measurements after twice the quarantine time $T_Q$ under total confinement (Hubei's measures), the death toll would follow the most probable law $D_T=7.9\times 10^4 \left(\log \left(\beta + \exp\left((t_{measures}-t_c)/T_L\right)\right)\right)^{\alpha}$. While this law matches Hubei's data and those of all countries having enacted measures less than 6 days after $t_c$, it is an extrapolation based on our model.

\section*{Dynamic self-similarity as an alternative to other current approaches}

Deterministic and stochastic compartmental models (SIR and SEIR) for infectious diseases have been used for epidemiology for COVID-19 \cite{Tang2020,Peng2020}. These models assume that the population is uniform and homogeneously mixing. Deterministic models constitute a Cauchy problem governed by a set Ordinary Differential Equations (ODEs) with constant coefficients. Also, a stochastic framework could be considered in these ODEs to model a more realistic infection problem during the early stages, when the number of infected individuals is small \cite{Liu2016,Iwata2020}. Mixing depending on the population age and all other factors can be  taken into account based on different contact geographic rates and social-economic groups. Non-uniform spatial distributions can also be analyzed by using networks connecting the different nodes into which compartmentalized populations evolve by means of SIR or SEIR models
\cite{Khaleque2020,Biswas2020}.

We propose a drastic departure from those approaches and make use of the laws of big numbers that this problem may follow. To do this, as a first approximation we assume a statistical homogeneity of the medium (human ecosystem) where the process of infection, incubation and possible mortality of COVID-19 develops into a pandemic. This allows a dramatic reduction of the relevant independent variables to time only, and supports the possible existence of an asymptotic power-law (a scaling law) in this variable. Two non-dimensional parameters will encapsulate: (i) the fundamental properties that the medium exposes to the action of the virus, and (ii) a simple model for the early behavior of the system prior to the asymptotic regime. The analysis of data shows that a surprisingly homogeneous and universally valid power law can be found for two rather general conditions found worldwide in this specific pandemic: (i) when a country is not prepared for the pandemic and (ii) containment measures are enacted with a delay. A few days of delay from the moment that a few cases of infection (about 100) are confirmed are sufficient to observe what we call a {\it free expansion}. In addition, the early behavior is shown to be relevant only for very small times compared to the time of infection $T_L$, which in this work we show it is 20.1 days in average. Based on the accuracy of the simple model obtained when compared to measurements of data for both confirmed cases of infection and deaths, the consistency of our approach is sustained.

\section*{Dimensional analysis of the time series}

The usual representations of the pandemic evolution in the media focus on the values of $C(t)$ and $D(t)$. Drastic confinement strategies like the one followed by China have proven reasonably valid to contain the propagation of the pandemic, while the recent trend in Korea points in the direction of a selective isolation (releasing non-infected population from confinement) as a even more effective strategy, showing a contained death toll while being friendlier to the economy of the country. 

The time series are usually represented in linear-linear scales\cite{Cheng2020}. This representation hides crucial dynamical behaviors when the values are still low. This precludes the identification of crucial time shifts, the incubation time, or a good estimation of the actual infection rate to match potentially universal behaviors among the countries. Current knowledge of this critical matter, which fundamentally impacts the country's sanitary and human resources, emphasizes dramatically the importance of an adequate knowledge and a statistical predictor to take the most appropriate measures and to gauge the death toll associated to the pandemic. 


\section*{Simplifying hypotheses}

First, $C(t)$ can be considered a statistical descriptor of the total population $S(t)$ with the capacity to infect others at a given time. It is reasonable to think that the majority of tests have been made to those persons developing symptoms, presumably an approximately constant fraction (with a homogeneously statistical distribution) of the total infected population $S(t)$. Indeed, one may hypothesize that human species reacts to this novel virus in a nearly homogeneous way worldwide. The consistency of the model with measurements is a valid test of this hypothesis. Thus, $D(t)$ results a crucial estimator to take early measures to contain the expansion of the pandemic. Moreover, $D(t)$ is a robust quantifier of the pandemic's toll. We show how $D(t)$ follows the same laws of $C(t)$, which, again, provides additional support to the hypotheses here made.

To study the dependencies of both $C(t)$ and $D(t)$ with the multitude of parameters involved, we seek for simplifying hypotheses to drastically reduce the complexity of the analysis. In what follows we show that the universe of parameters can be reduced to a minimum, which reduces the number of degrees of freedom of the system without compromising the accuracy of predictions. To do this, we use the potential of a powerful tool, dimensional analysis, seeking for self-similarities which are usually reflected by the existence of power-law relationships.

\subsection*{Characteristic time}

The first fundamental independent variable of the problem is the time $t$. In this problem, given the circadian character of the records available, the time unit is a day. Among the possible choices, we have the average time of incubation, $T_I$, and the time $T_L$ during which a carrier of a viral load can infect others. While a self-similar expansion (a power-law behavior) would be independent of that choice, we need to rely on physical similarity with known phenomena, for example the exponential quenching of a homogeneous flame in a domain under total containment. This approach is compatible with SIR and SEIR models found in the recent literature for this phenomenon\cite{Liu2016,Cheng2020,Bittihn2020,Biswas2020,Iwata2020} when the appropriate leading characteristic times are identified. In this case, the time of infectivity $T_L$ should be equal to that of the exponential decay observed in the death rate during the spread of the pandemic in China provinces after total containment measures were taken. Following the procedures of dimensional analysis applied to available data, in this work we will obtain 20.1 days in average as the most plausible time of infectivity.

\subsection*{The human network preyed upon by SARS-CoV-2}

A pandemic develops due to three fundamental factors: (i) a number of infected people above a critical value (which depends on the infection rate), (ii) lack of prior immunity of the population, and (iii) social unpreparedness. These are the factors that make human population a ``good combustible'' for the spreading of a virus: when an opportunistic pathogen finds its way to infect unprepared people and produce a disease in a sufficient number of individuals, its evolution is akin to that of the expansion of a flame in a combustible. The mechanisms by which the flame expands in space are intrinsically related to its dimensional character. In each dimension, the flame propagates at a constant speed (i.e., the space traveled would be proportional to $t$) whenever it finds a combustible with homogeneous properties along its way. For example, the area of a forest burned by a wildfire is proportional to $t^2$, while the volume burned by a spark in a homogeneous mixture of fuel and oxidant is proportional to $t^3$. However, when humans are exposed to an initially localized pathogen (e.g. the SARS-CoV-2), human society can be considered a nearly homogeneous network (which does {it not} imply a spatial homogeneity) for the propagation of a signal like the virus infection. To do this, we assume that the majority of the infection process takes place in cities, where human population is concentrated. In reality, humans behave roughly similarly in cities, which is the most populated geographical form of current human ecosystems, nearly independently of the culture. Moreover, humans interact with others at a rate that can be considered much more homogeneous than the spatial concentration of population itself. In fact, among other activities, humans move in ample spaces, seek for food from and visit toilets on a relatively homogeneous daily basis, either sharing fluids or leaving biological residues around them at the immediate reach of others in solid surfaces, in liquids, and airborne (aerosols). In other words, the human network is not a three dimensional volume of more or less static points (persons), but a relatively homogeneous fractal network where a physical signal can propagate following principles of self similarity\cite{Song2005,Serrano2008}. Depending on the ability of the pathogen to find receptors in the host (the ease of contagion), this macroscopic network may exhibit different fractal dimensions. Those fractal dimensions are expected to be larger than 3.

In summary, before governments enact measures to control the pandemic, COVID-19 propagates in an unprepared population which exposes a virgin network to the virus. We hypothesize that the human network has a fractal dimension that should be nearly homogeneous in all countries having people living in cities.

\subsection*{Self-similar infection process}

Continuing with the physical analogy, a real flame never develops in a self-similar way (e.g. spherically) from the very beginning.  there is a characteristic size of the population with the sufficient viral load to trigger a free expansion using the existing (statistically homogeneous, in average) human network. In the absence of containment measures, when the expansion evolves with times of the order of the infection time, the expansion (number of infected people with time) should exhibit a self-similar behavior. Since the real size of the infected population is unknown, we should rely on self-similarity arguments. If both $C(t)$ and $D(t)$ present a power law behavior, then the actual population can be assumed to behave in the same way.

At this point, we can simply make $C(t)$ non-dimensional with any tentative value of $C_C$. If the typical policy of all countries has been to take tests to anyone presenting symptoms and seeking for remedy at a hospital, $C_C$ can be reasonably hypothesized to be constant for all countries. We observe that, after the appropriate identification of the beginning of the expansion, the data for those countries with the largest expansion rates (i.e. those which suffer the real pandemic) collapse independently of the value of $C_C$.

\subsection*{The geographic expansion: initial infection time of a country}

Current measures of the pandemic expansion are performed in fixed geographical areas, i.e. countries and, in some cases, provinces. One may assume that the size of those areas hold a sufficiently large population to exhibit a homogeneous, large network, fundamentally concentrated in cities. The pandemic was originated in a certain city (Wuhan, China). Naturally, the required time to develop a critical size from the onset of the infection at Wuhan varies from country to country depending on its own connectivity with the location of the origin of the pandemic. Under the assumption that a self-similar behavior (presumably, a power-law dependence with time) develops in a given country, the delay times $t_c$ should be calculated as those which collapse the measured values (i.e., both $\varphi_c=C(t-t_c)/C_C$ and $\varphi_d=C(t-t_c)/(m C_C)$) and a self-similar behaviour among different countries may emerge.

\subsection*{Raw data}

Raw data for $C(t)$ and $D(t)$ are available at \url{https://github.com/CSSEGISandData/COVID-19/tree/master/csse_covid_19_data/csse_covid_19_time_series} in the form of daily time records of 238 countries and provinces since January 22, 2020.

\section*{Analysis to verify hypothesis and calculation of main variables}

\subsection*{Analysis of confirmed cases $C(t)$}

Here, we assume the hypothesis that $C(t)$ can be considered as a representative descriptor of the infected population $S(t)$, while $D(t)$ is a robust descriptor of the death toll of the pandemic.

We first find the values of $t_c$ which produce the best collapse of $C(t)$ to an expected behavior of the form $\tau^\alpha$. To do this, we select a set of five complete series of countries with these features: (i) the corresponding governments enacted measures a number of days after their ``zero'' $t_c$, and (ii) the infection has developed from relatively few and localized foci (ideally, just one) with a sufficient viral load. We have selected the following countries whose series seem to have those features: Italy, Hubei (China), Spain (the majority of cases concentrated in Madrid), Germany, South Korea, and Sweden. For comparison, we also include in the study two regions where measures were enacted in the whole country {\it before} their {\it local} $t_c$ was observed. Even in these cases, there is a net growth of the infection due to an unavoidable initial infection load, which decays exponentially (this is well reflected by SIR and SEIR models\cite{Liu2016,Iwata2020}) in times of the order of or smaller than the characteristic average time of infection.

\begin{figure}[hb]
\begin{center}
\includegraphics[width=0.95\textwidth]{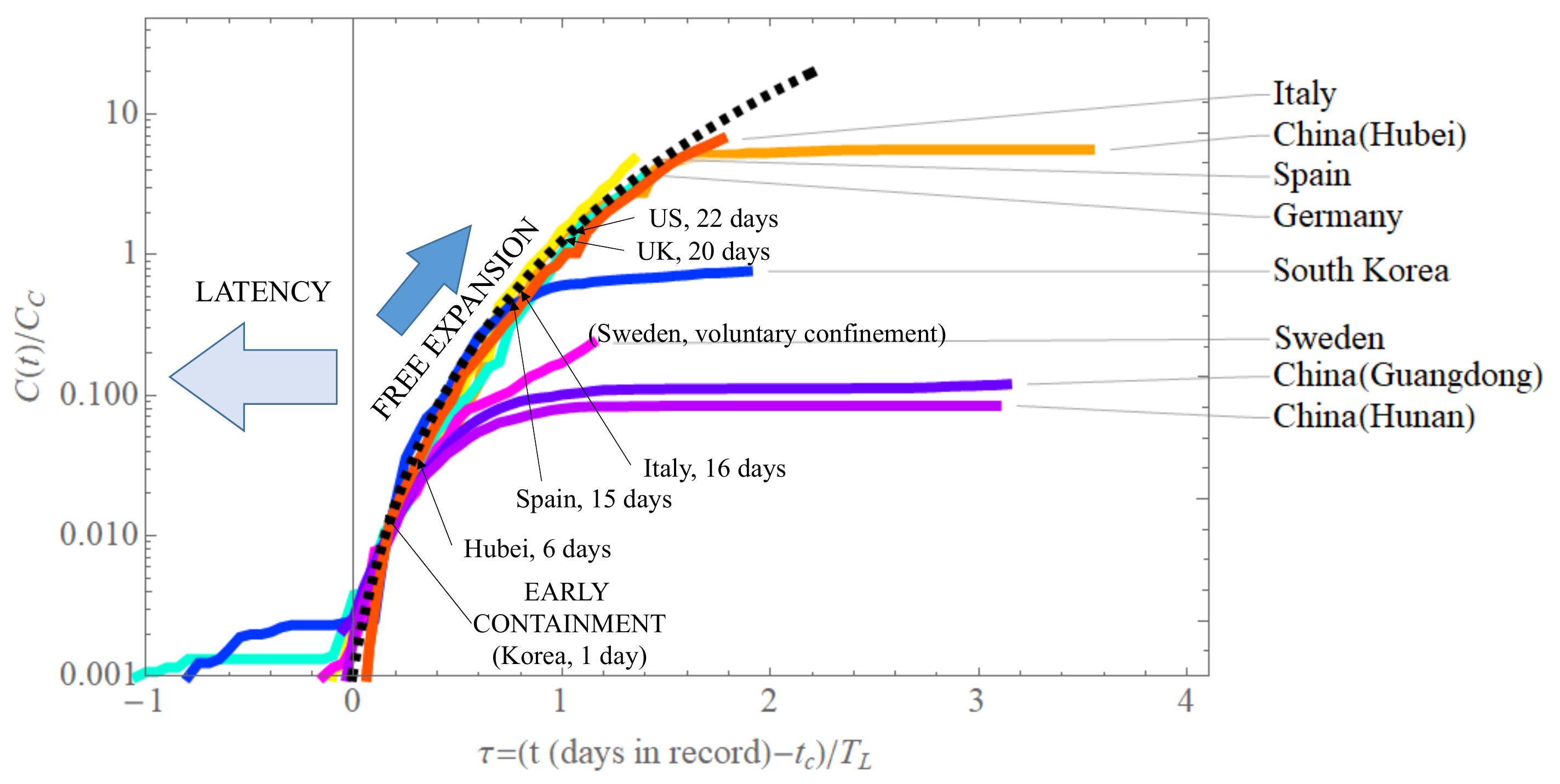}
\end{center}
\caption{The non dimensional collapse of Confirmed series ($C(t)$). Seven selected countries exhibiting free expansion are selected to illustrate the effect of measures. The dashed line is the function $\varphi=(\log\left(\beta+\exp(\tau)\right)^\alpha$, with $\alpha=3.75$ in this study. In this plot, we have used the final values obtained from the complete analysis, i.e. $C_C=1.2 \times 10^4 $ and $T_L=20.1$ days. The points where different countries took containment measures are indicated, together with the corresponding delay times in days.}
\label{fig1}
\end{figure}

We make a classical search for the values of $t_c$ which make the series collapse with a reference series. To do that, we take as the first reference series that from the country where the pandemic initiated (Hubei), and set the origin of its expansion as $t_c=0$. We observe that such a collapse is easily found (see figure \ref{fig1}). At this point, we readily observe that the collapsed data of the most impacted countries by the pandemic follow a power law behavior as $\varphi_c\equiv C(t-t_c)/C_C=\left((t-t_c)/T_L\right)^\alpha\equiv \tau^\alpha$ during long time periods (at least before containment measures are enacted) where the choice of $C_C$ and $T_L$ is arbitrary at this point. This leads to an average value of $\alpha\simeq 3.75\pm 0.05$. In figure \ref{fig1}, we represent the series $C(t)$, where the value of $C_C$ is made equal to that which leads to the simplest form of the power law at times of the order of unity, i.e. $\varphi_c\simeq \tau^\alpha$, once the characteristic time $T_L$ has been calculated from available data.

\subsection*{Mortality analysis, $D(t)$}

Then, we do the same approach for the deaths $D(t)$ time series, but this time we use $C_C$ multiplied by a mortality $m$ such that $\varphi_d=D(t-t_d)/(m C_C)$ would commensurate with $\varphi_c$. We use the time series of five representative countries with the highest growth rate and larger death tolls (Italy, Hubei, Spain, Iran, and UK). Using the same approach as before, we seek for the best collapse and the minimal global differences with respect to a power law $\tau^\alpha$, with $\tau=(t-t_d)/T_L$ and the same $\alpha=3.75$. This gives us the values of $t_d$. We observe that the values of $m$ producing a prefactor of the order unity can be considered homogeneous, an observation supported by including Netherlands, Germany, Belgium and Switzerland as additional representatives of free expansion. Asian countries like Korea and two Chinese provinces are also included to illustrate the effect of containment.

\begin{figure}[ht]
\begin{center}
\includegraphics[width=0.95\textwidth]{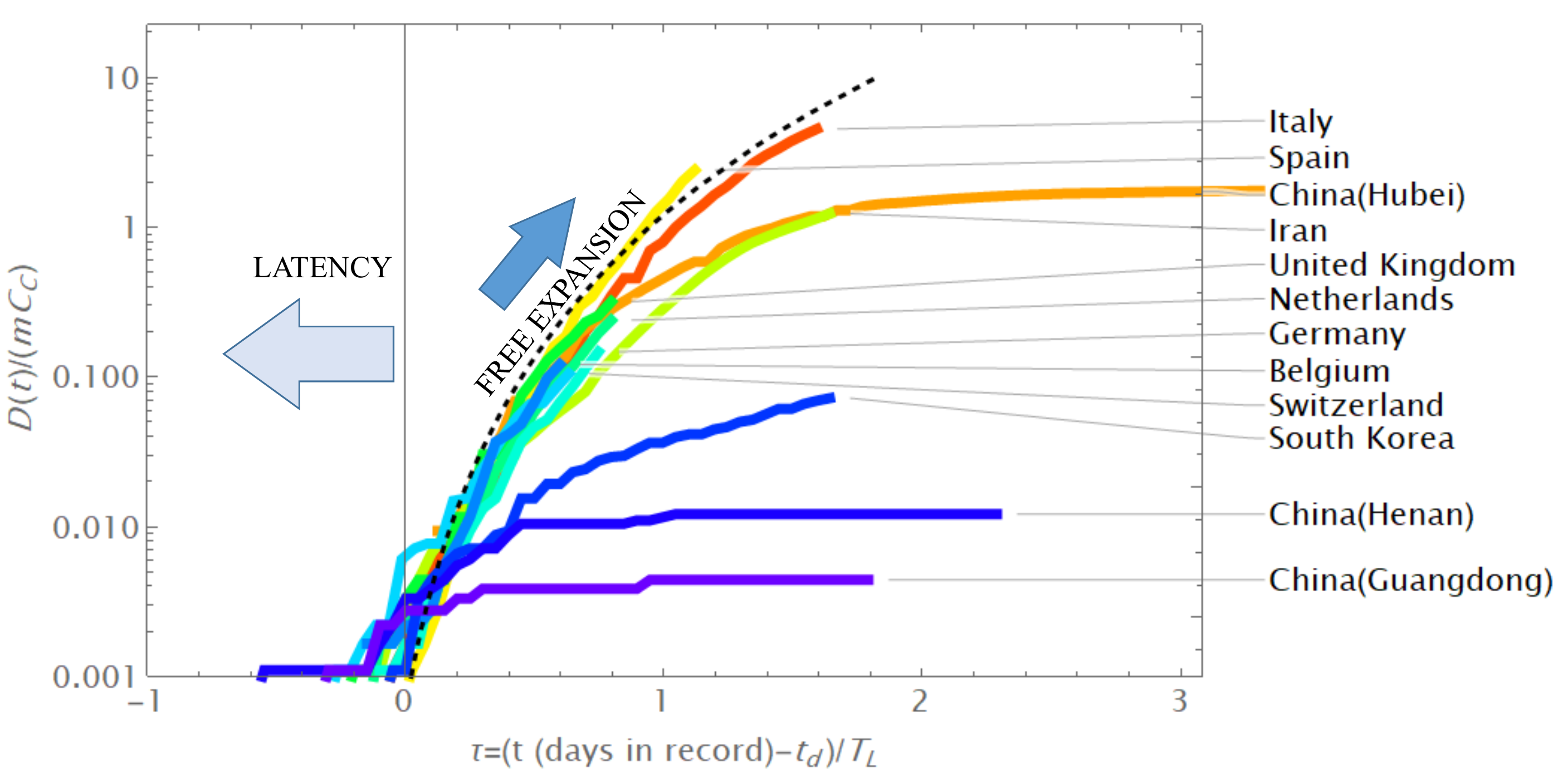}
\end{center}
\caption{The non dimensional collapse of Deaths series, $\varphi_d=D(t-t_d)/(m C_C)$. 11 countries are selected. The dashed line is the {\it same\rm} function as for $\varphi_c=C(t-t_d)/C_C$}
\label{fig2}
\end{figure}

This demands that both $C(t)$ and $D(t)$ should behave in the same way, thus obeying the same non-dimensional behavior once the appropriate delay times are found for each variable $C(t)$ and $D(t)$, and for each country. Measurements are plotted in figure \ref{fig2}.

\begin{figure}[ht]
\begin{center}
\includegraphics[width=0.95\textwidth]{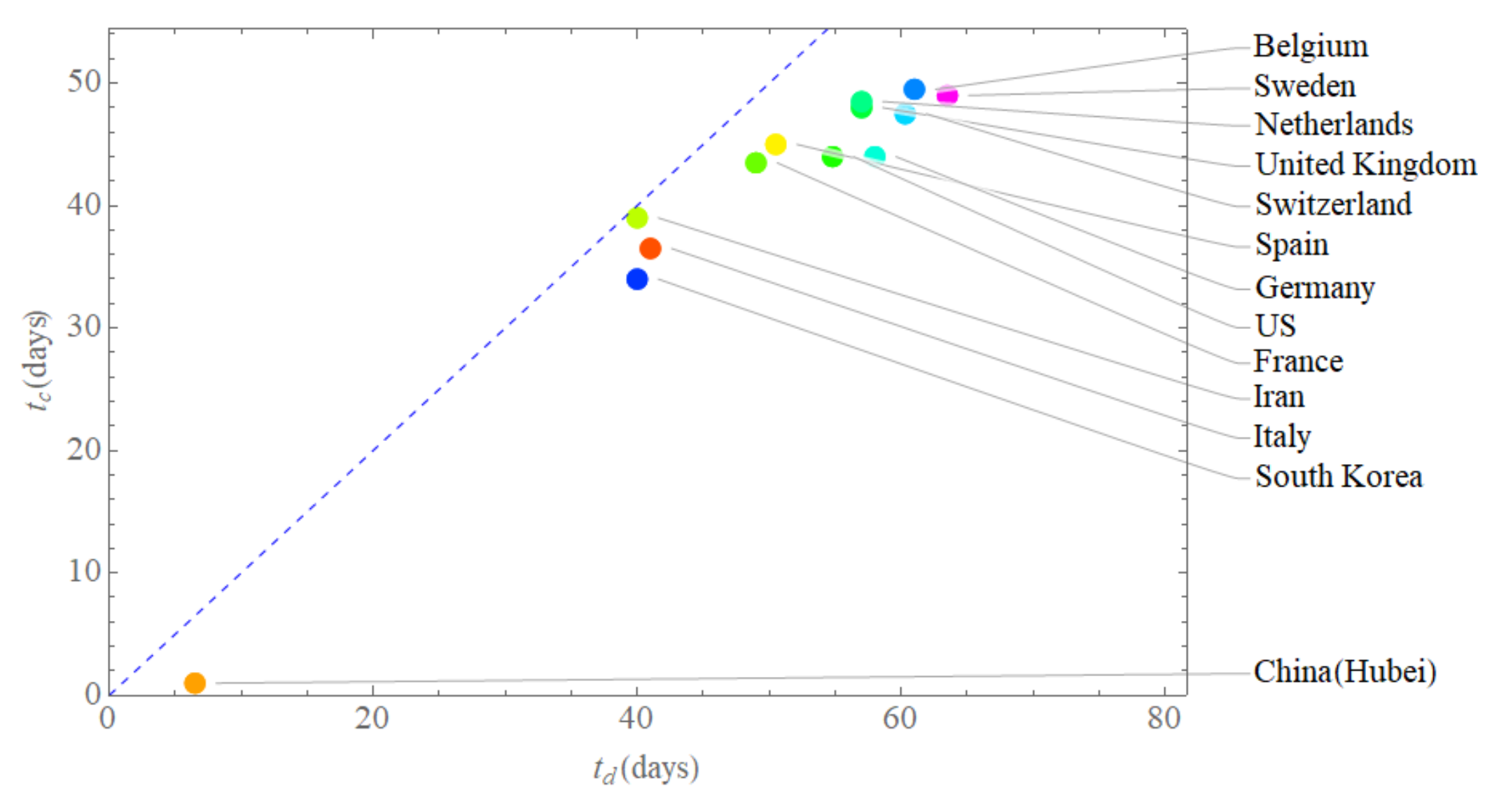}
\end{center}
\caption{Plot of the (dimensional) time delays $t_d$ for Deaths series (in abscissae) versus the time delays $t_d$ for Confirmed series. Blue dashed line is the identity line.}
\label{f3}
\end{figure}

The resulting mortality ratios $m$ should not be confused with the daily mortality rates $D(t)/C(t)$ since the delay times $t_d$ for $D(t)$ and $C(t)$ series {\it are different}.\astfootnote{We propose that $m$ represents a best estimator of the final mortality rate (i.e., $D(t_{final})/C(t_{final})$ when both $C(t) $ and $D(t)$ reach a plateau after measures to contain the pandemic are enacted. In fact, we hypothesize that $D(t_{final})/(m C(t_{final})=\gamma m$, where $\gamma>1$. Indeed, if we assume that the relative number of tests made (i.e., the number of tests made in relation with the size of the real infected population) increase linearly with time, then the final value of the mortality rate measured (i.e. $\gamma m$) will be about twice the average $m$ along the curve. Then we have $\gamma=2$ and $m=\left(D(t_{final})/(2 C(t_{final}))\right)^{1/2}$. The case of Hubei is probably the best for trying this hypothesis. The optimum value of $m$ for Hubei is around 0.15, while the final mortality rate $D(t_{final})/C(t_{final})$ has resulted 0.046. Then, $m=0.15$. The final asymptota of $D(t)/\left(m C(t)\right)$ results approximately equal to 0.28, which is indeed close to $2m=0.3$. This serves to additionally support the robustness of the hypotheses made to support the physics behind the model.}
In figure \ref{f3}, we plot the values of $t_c$ and $t_d$ found for our selection of countries with highest impact of COVID-19.

\subsection*{Time delay of confinement after free expansion. The characteristic time of infection $T_L$}

A fundamental observation that can be made upon the collapse of the time series is the existence of a close relationship between the delay between $t_c$ and the day $t_{measures}$ when containment measures are enacted, and the time when both $C(t)$ and $D(t)$ curves exhibit a final plateau. This secondary process is what we hypothesize as a {\it quenching} of the pandemic with an initial inertia that should exhibit a characteristic decay rate. The characteristic time of the problem, i.e. the time of infection, should necessarily surface in the observed behavior of both $C(t)$ and $D(t)$ in this secondary process. However, the robustness of mortality $D(t)$ as a direct descriptor of the products of the pandemic (i.e. the "combustion") suggest to work with these series. Thus, we represent $y=(t_{90\%}-t_{measures})/T_0$ versus $\Delta \tau=(t_{measures}-t_d)/T_0$ and seek for its possible relationship, where $t_{90\%}$ a representative final time: the time value when 90\% of the long term death toll is reached in the series. Fortunately, we have complete series with a sufficiently long asymptota to al lest make linear estimations. The value of $T_0$ which makes the data fit to the simplest form $y=1+A \Delta \tau $ is $T_0=20.1$. Remarkably, measurements yield $A=1$ taking this value of $T_0$ too. Thus, we take this as the time of infection $T_L$ used in the whole analysis.

\begin{figure}[ht]
\begin{center}
\includegraphics[width=0.75\textwidth]{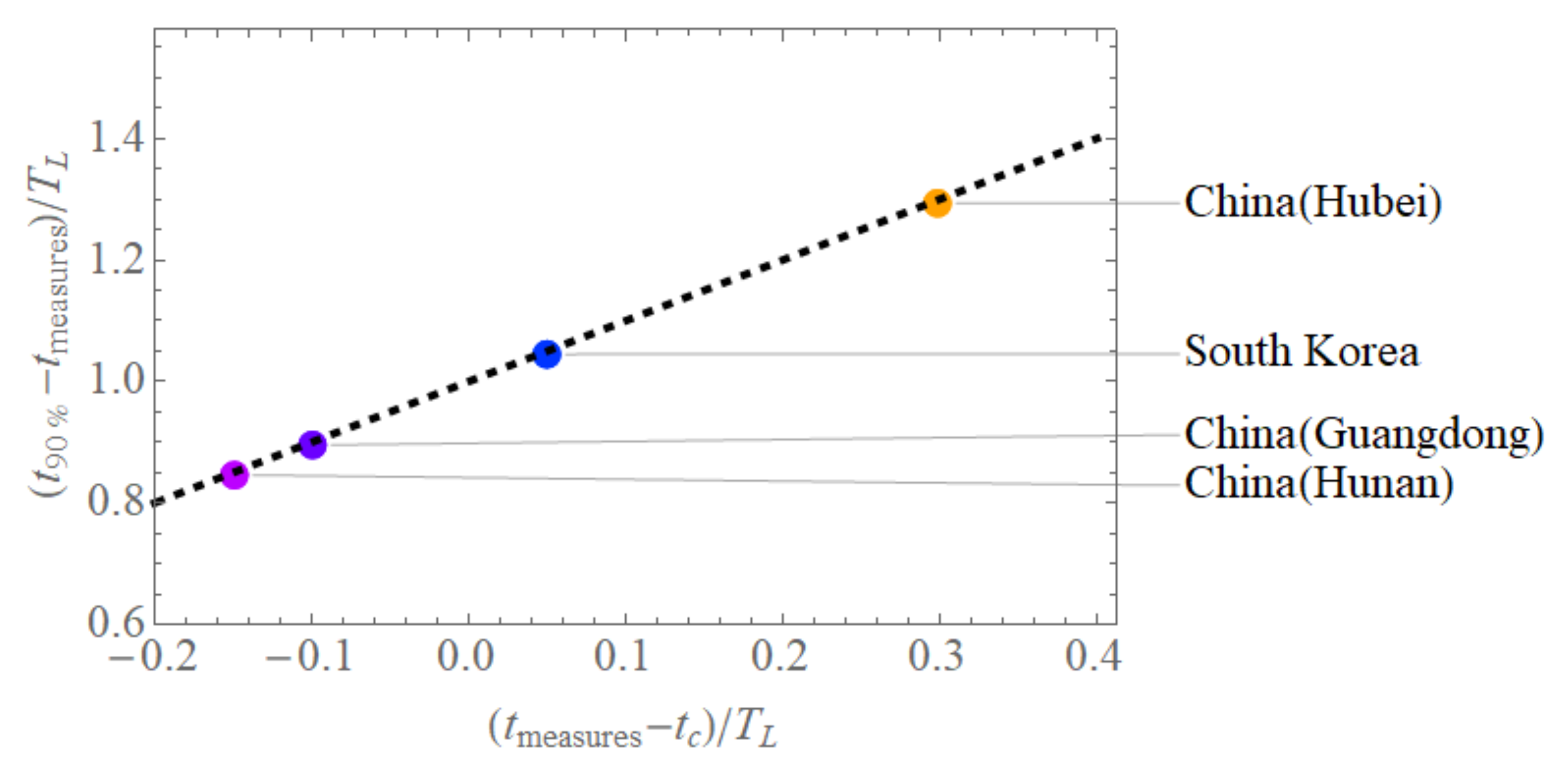}
\end{center}
\caption{Plot of the non dimensional time $(t_{90\%}-t_{measures})/T_L$ versus the non dimensional time delay $\Delta \tau=(t_{measures}-t_d)/T_L$ from the initiation of the free expansion up to the date when measures were taken. Here, $T_{90\%}$ is the time required to reach a 90\% of the plateau from the day when total containment measures were taken. $T_L$ is made equal to 20.1 days to get $y=1 + \Delta \tau$}
\label{f4}
\end{figure}

In the cases where the delay is negative (i.e., no free expansion), the origin has been taken at the point where maximum growth rate is observed. This powerful result leads us to propose a simple formula for a minimum quarantine: if we wanted to reach a situation where approximately 90\% of the expected mortality of an indefinite confinement is reached, we should keep confinement this number of days:
\begin{equation}
T_Q = T_L\left(1+(t_{measures}-t_c)/T_L\right)\equiv T_L\left(1+ \Delta \tau \right)
\end{equation}
Obviously, since total containment could wreak havoc in a country's economy for delays over 15 days (just a few days of delay is critical when power laws are involved), alternative measures (e.g. selective confinement) would become strictly necessary.

Once a best characteristic value of $T_L$ has been found, one can adjust the value of $C_C$ to produce a simple relationship as $C(t)/C_C=\tau^\alpha$, which is equivalent to calculate the value of the dimensional prefactors which multiplied by the function $\varphi=\left(\beta+\exp\left(\tau\right)\right)^\alpha$ fit both $C(t)$ and $D(t)$ time series according to previous analyses.

\subsection*{Delays in taking measures for the different countries. Predicted associated death tolls.}

The observed accuracy of our self-similarity model to fit both the data of confirmed cases and the deaths supports the hypothesized statistically homogeneity of main properties of the human network worldwide. Taking advantage of this, our self similarity model leads to a simple way to calculate the most probable death toll at the end of a sufficiently long confinement after measures are enacted. Since final mortality would be proportional to the true population of infected people at the time of enacting measures, our previously validated hypotheses also support that the final mortality under containment will be proportional to $C(t_{measures}-t_c)$. Fitting the data of the global experiment of China, our model yields the following formula for the expected {\it average} deaths toll $D_T$:

\begin{equation}
D_T=8\times 10^4 \left(\log \left(\beta + \exp\left((t_{measures}-t_c)/T_L\right)\right)\right)^{\alpha}
\label{death}
\end{equation}

\begin{figure}[ht]
\begin{center}
\includegraphics[width=0.75\textwidth]{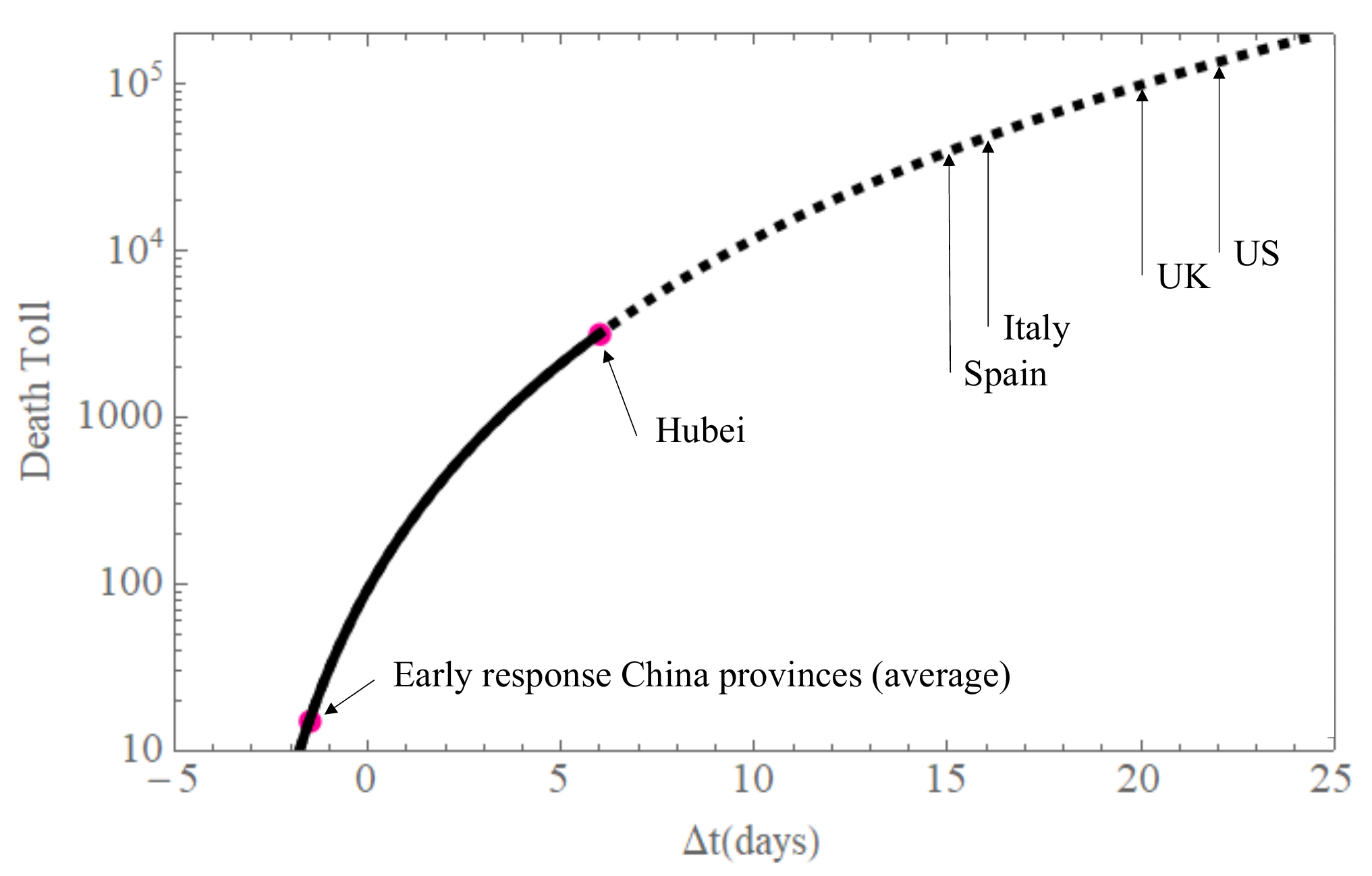}
\end{center}
\caption{Death toll as a function of the delay time $\Delta T=t_{measures}-t_d$. The points in time where different countries have taken measures are indicated.}
\label{f6}
\end{figure}

Thus, according to these data and the validity of the hypothesis of statistical homogeneity, the prediction \ref{death} would be valid in any urban region of the world when total confinement measures (the ones taken in Hubei) are enacted at the moment when more than 100 confirmed cases are observed. This critical value $C=100$ has been obtained by averaging the values of $C(t)$ where the largest second derivative is approximately observed (i.e. the sudden raise of the free expansion) in figures \ref{fig1} and \ref{fig2}). Here, $\alpha=3.75$ according to the global model for both $C(t)$ and $D(t)$. The fitting to the early time measures of average Chinese provinces yields $\beta\simeq 0.23$. This is plotted in figure \ref{f6}.

We summarize in Table 1 some of the results upon which our model and predictions are based, and some extrapolations for selected countries. Mortalities are statistical estimations with a typical variability as the one that can be observed among death curves $\varphi_d$ (e.g. Spain and Italy, see figure \ref{fig2}).
\begin{table}[hb]
\begin{tabular}{@{\vrule height 10.5pt depth4pt  width0pt}cccccc}
Country & Time delay & Death toll & Est. quarantine & Real quarantine \\
 & (days) & (est. or real) & (2$\times T_{90\%}$ days) & enacted (days) \\ \hline
\hline
Hubei (China) & 6 & 3160 & 54 & 63 \\
\hline
Spain & 15 & 38950 & 70 & -- \\
\hline
Italy & 16 & 47700 & 72 & -- \\
\hline
UK & 20 & 98100 & 80 & -- \\
\hline
US & 22 & 134800 & 84 & -- \\
\hline
\end{tabular}
\caption{Delay times in enacting measures at some selected countries and regions with respect to the origin of the free expansion of the pandemic.}
\end{table}

We are especially grateful to Cristina de Lorenzo for very fruitful discussions. Also, Pascual Riesco-Chueca gave us important suggestions. Our recognition to Juan Antonio Guti{\'e}rrez del Pozo for his suggestions and for having warned A.M.G.C. as early as February 10th, 2020 on the magnitude of the problem that Spain would face under this pandemic.


\end{document}